# A Study on Impacts of RTT Inaccuracy on Dynamic Bandwidth Allocation in PON and Solution

Son Nguyen Hong[1], Hao Nguyen Anh[2], and Thua Huynh Trong[3]

[1,2,3]Department of Information and Communication Technology, Post and Telecommunication Institute of Technology, Ho Chi Minh City, Viet Nam

## ABSTRACT

*The circle travelling delay between OLT (Optical Line Terminal) and ONU (Optical Network Unit) is one of most important items in dynamic bandwidth allocation (DBA) algorithms in PON, called RTT (Round Trip Time). The RTT is taken into account when OLT assigns the start times for upstream bandwidth grants. In most case, RTT is estimated before making bandwidth allocation decisions in dynamic bandwidth allocation algorithms. If the estimated RTT is incorrect, the bandwidth allocation decisions are not matched with bandwidth requests of channels. Thus, performance of PON can get worse by deviation of RTT. There are several reasons that cause the RTT to be varying, such as processing delay, distance of OLT and ONU, changing in fiber refractive index resulting from temperature drift, and degree of accuracy of RTT estimation methods. In this paper, we evaluate the impacts of RTT inaccuracy on performance of DBA and identify levels of collision and waste of bandwidth. By this way, we propose a method to remedy the performance degradation encountered by the situation.*

## 1. INTRODUCTION

Nowadays, the growth in traffic flowing through networks continues to increase due to the spread of numerous applications and services that are characterized by large volumes of content and massive amounts of data traffic. As a network infrastructure to support such huge traffic, PONs (Passive Optical Networks) are widely deployed in the access networks. Various PON standards have been developed, including ATM-PON (Asynchronous Transfer Mode-PON), BPON (Broadband PON), GPON (Gigabit PON), LR-PON (Long-Reach PON), SFiWi (Smart Fiber-Wireless) and EPON (Ethernet PON). The EPON is used as a good solution for ISPs to offer faster Internet connectivity so as to keep with subscriber demands. Although, several aspects of the PON standards need to be improved, such as dynamic bandwidth allocation, synchronization, performance, energy consumption, etc. The loss of synchronization between OLT and ONUs is a common problem that results in degradation of network performance. Using an incorrect RTT value in algorithms of PON is one of reasons that lead to the problem.

RTT is period of time that is from OLT starting to signal for ONU until OLT receiving returned signal from the ONU. Since OLT allocates bandwidth to ONUs according to TDM scheme, dynamic bandwidth allocation algorithms (DBAs) always include RTT to make sure that the allocation is fair, low delay and high utilization.

The RTT is a factor that is carefully considered in any solutions related to PON. In case of LR-PON that has distance of transmission to reach 100 km, RTT increases up to 1 millisecond. In order to combat the detrimental effect of the increased RTT, the research in [1] proposed the Multi-Thread Polling algorithm, that runs several polling threads in parallel, and each of the threads is compatible with the proposed DBA algorithms in traditional PON. The work in [2] proposed a novel online Just-in-Time (JIT) scheduling framework that gives the OLT more opportunity to make better scheduling decisions than standard online scheduling. In this framework they used the largest RTT in the EPON for timing of the GATE message transmissions to ensure that any ONU receives the GATE message in time. In [3] about cutting down the wasteful energy consumption in SFiWi networks, they carefully take into account the RTT when determining the ONU sleep state period. They also specified that the optimal sleep state periods vary depending on the propagation delay (RTT) in wireless networks.

According to [4][5][6], OLT sends periodic timestamps to ONUs in order to correct their clock for drift error. Since RTT from OLT to various ONUs can vary over a large range of values, OLT has to send individually adjusted timestamps to guarantee the correctness of the timestamp when it arrives at the ONU. This requires the OLT to know the RTT to every individual ONU. Lack of this information, the OLT can not arbitrate the upstream channel in a collision free manner. As described in [5][7], to calculate time slots for ONUs, OLT has to know information of ONUs, such as operation state, whether sending data or not, what kind of service, whether or not continue to receive data from users, how much remaining data in ONUs, etc. OLT is up to date with the information by ONU's response messages. The responses must be received at exactly time by OLT in order to allocate time slots efficiently.

The dynamic bandwidth allocation algorithms in OLT take the role of time slot calculation for each ONUs. As important factors in bandwidth allocation, the time of starting and the time of finish a transmission from each ONUs to OLT must be exactly. The efficiency of bandwidth allocation algorithms depend not only on method of calculating long size of time slot but also on the points of time. If an ONU transmission is soon started while previous next ONU transmission has not finished yet, a collision occurs and the transmission fails. Otherwise, too late starting next ONU transmission will result in waste of bandwidth of the system.

If the value of RTT known by ONU differs from its right value, OLT can not be exactly updated ONU's information due to wrong point of time in response from ONU. DBA algorithms can not exactly allocate time slots for ONUs without accurate information about ONUs. The collisions or waste of transmission time may occur, that lead to performance degradation in PON.

In fact, some deviation of new RTT from the previously measured RTT may be caused by changes in fiber refractive index resulting from temperature drift. Additionally, clock drift results in RTT inaccuracy, which causes OLT and ONUs not remain in synchronization for long periods of time. There are other reasons that also lead to RTT inaccuracy in PON. Analyzing impacts of the inaccuracy on performance of DBA is necessary for finding solutions to improve the performance in PON networks. However, there are no researches of the issue in details.

In this paper, we, firstly, analyze cases of RTT inaccuracy and consider how PON system getting worse with it. Then, we propose a method for mitigating the degradation of DBA performance encountered by the RTT inaccuracy. The rest of paper is organized as follows: Section 2 describes the mathematic model of OLT's scheduling scheme and analyzes impacts of RTT inaccuracy on DBA performance. The impacts are evaluated by simulations in Section 3. Our proposal is presented in Section 4, and it is validated by results of simulation in this section. Finally, the paper is finished by conclusions in Section 5.

## 2. MODELLING THE ONUs GRANT SCHEDULING SCHEME OF OLT

Considering an EPON system includes an OLT and *n* ONU. In cycle *j*, each ONUs has a data transmission time *l(i, j)*. The values *l(i, j), RTT(i)* have been updated before OLT calculates the allocated bandwidth for ONUs. IPACT algorithm is applied to this model to evaluate the influences of RTT inaccuracy.

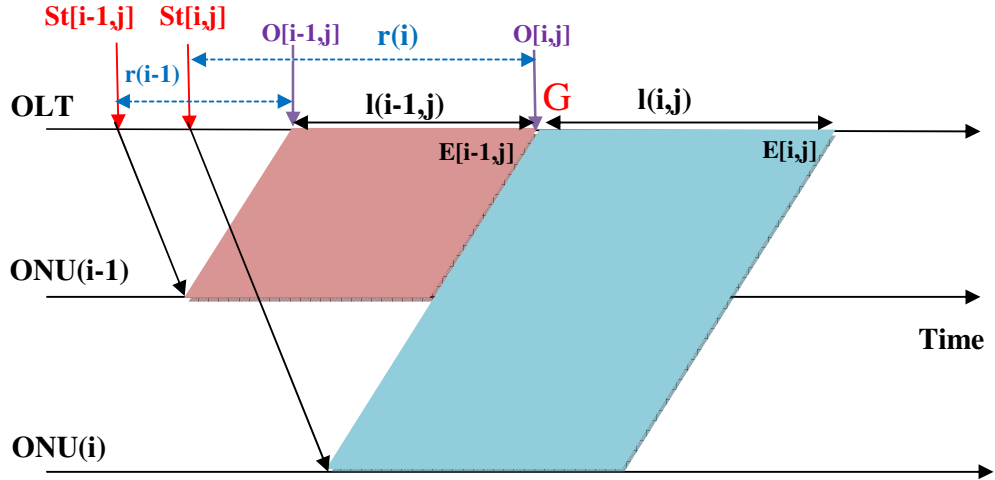

Figure 1. Operation Model of OLT and two ONUs

The figure 1 describes the data transfer operation between the OLT and two adjacent ONUs (ONU(i) and ONU(i-1)) in cycle j, which relates the following expressions and parameters:

- St [i, j]:  staring time of signal transmission for ONU( i) in cycle j.
- St [i-1, j]:  staring time of signal transmission for ONU (i-1) in cycle j.
- r(i): RTT of  ONU( i).
- r(i-1): RTT of  ONU (i-1).
- l(i,j): the length of data transfer time for ONU( i) in cycle j.
- l(i-1,j): the length of data transfer time for ONU(i-1) in cycle j.
- O[i-1,j] : the time that OLT opens gate for ONU(i-1) sending data in cycle j.
- O[i,j] : the time that OLT opens gate for ONU(i) sending data in cycle j.

$$O[i-1,j] = S[i-1,j] + r(i-1)$$
$$O[i,j] = S[i,j] + r(i)$$

- E[i-1,j]: the time that OLT closes the sending gate of ONU( i-1) in cycle j.
- E[i,j]: the time that OLT closes the sending gate of ONU( i ) in cycle j.

$$E[i-1,j] = St[i-1,j] + r(i-1,j) + l(i-1,j)$$
$$E[i,j] = St[i,j] + r(i,j) + l(i,j)$$

Theoretically, all bandwidth allocation algorithms consider that the starting time of ONU(i) coincides with the ending time of ONU(i-1), this point is G in figure 1. Thus,

$$E[i-1,j] = O[i,j]$$

At the time of G, OLT will be updated the followings:

- The time that OLT should send granted message to ONU(i-1)
- The length of transmission time for ONU(i-1)
- The time that OLT should send granted message to ONU(i)
- The length of transmission time for ONU(i-1)

The RTT of ONU(i-1) and RTT of ONU(i) are still kept as those of previous estimation, no update processing for them. Thus, the information may differ a certain $\Delta x$ from the actual values and the start time of ONU(i) may differ a certain $\Delta y$ from the calculated O[i,j]. The stop time of ONU(i-1) also differ a certain $\Delta y'$ from its calculated value. In fact, expressions are as follows:

$$E[i-1,j] = St[i-1,j] + r'(i-1,j) + l(i-1,j).$$

$$O[i,j] = St[i,j] + r'(i).$$

Let $\qquad \Delta t = O[i,j] - E[i-1,j],$

hence $\qquad \Delta t = St[i,j] + r'(i) - St[i-1,j] - r'(i-1,j) - l(i-1,j).$

The operation performance of DBAs depends on value of $\Delta t$ as follows:

- If $\Delta t > 0$, which causes waste of bandwidth

- If $\Delta t < 0$, which causes conflicts (overlap of periods of time).

- If $\Delta t = 0$, this is ideal, all are the same as results of calculations

To keep $\Delta t = 0$, either DBAs make sure that calculated RTT never differs from its actual value or we have a solution for combating the RTT deviation. It is difficult for DBAs to guarantee exactly RTT. This is the reason why we look for other solution.

In conditions of RTT inaccuracy, DBAs should take account of trade off between collision rate and channel utilization in response to real demands.

Let $k$ is number of successful ONUs in data transmission; $n$ is number of ONUs in system; $n$-$k$ is number of failed ONUs. The collision rate is calculated by

$$R_{col} = \frac{n-k}{n}$$

The total of waste of transmission time in a cycle is calculated by

$$W = \sum_{i=1}^{k} \Delta t(i)$$

The total of time that all ONUs spent for sending data to OLT in a cycle is calculated by

$$H = \sum_{i=1}^{k} l(i)$$

The line utilization is calculated by

$$U = \frac{H}{(H + W)}$$

## 3. SIMULATION AND EVALUATION

In this section, we evaluate performance of the PON mechanism according to cases of RTT inaccuracy. In our simulation program, the value of RTT deviation is randomly taken from the range of $[-\Delta x; + \Delta x]$. The simulation program is developed as a PON system with configuration of 64 ONUs. The collected parameters include the collision rate, the waste of transmission time and the line utilization.

Case 1: with 64 ONUs, the deviation of RTT is allowed in range of [-1,1]. The results of 20 successive cycles of simulation are shown in table 1.

Table 1. Results of the case of $\Delta x = \pm 1$

| Cycle | ONU | Collision Rate | Waste of Trans Time($\mu$s) | Line Utilization |
|-------|-----|----------------|------------------------------|------------------|
| 1  | 64 | 50.00 | 17.14 | 91.95 |
| 2  | 64 | 53.12 | 17.98 | 91.75 |
| 3  | 64 | 48.43 | 18.79 | 88.71 |
| 4  | 64 | 50.00 | 18.80 | 91.78 |
| 5  | 64 | 48.43 | 18.96 | 91.89 |
| 6  | 64 | 50.00 | 18.97 | 90.57 |
| 7  | 64 | 50.00 | 19.23 | 91.18 |
| 8  | 64 | 42.18 | 19.55 | 91.45 |
| 9  | 64 | 51.56 | 19.64 | 91.48 |
| 10 | 64 | 50.00 | 20.09 | 90.95 |
| 11 | 64 | 46.87 | 20.19 | 91.83 |
| 12 | 64 | 53.12 | 20.44 | 92.31 |
| 13 | 64 | 48.43 | 20.89 | 91.29 |
| 14 | 64 | 51.56 | 21.50 | 90.59 |
| 15 | 64 | 46.87 | 21.72 | 90.97 |
| 16 | 64 | 46.87 | 22.36 | 88.26 |
| 17 | 64 | 46.87 | 23.24 | 90.85 |
| 18 | 64 | 46.87 | 23.53 | 90.13 |
| 19 | 64 | 45.31 | 23.82 | 90.30 |
| 20 | 64 | 43.75 | 25.69 | 90.29 |

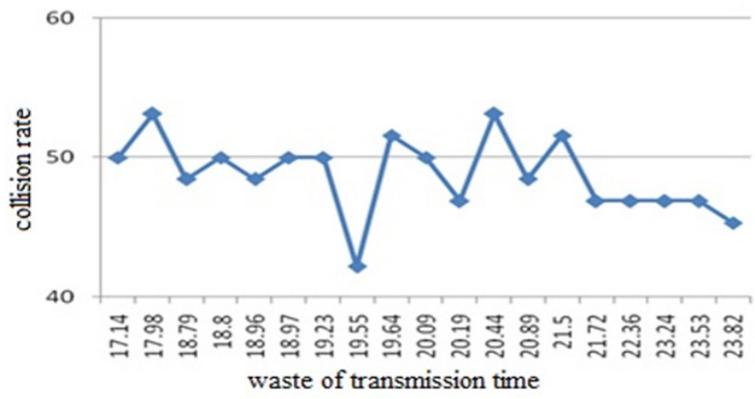

Figure 2. Relationship between collision rate and waste of transmission time with Δx = ± 1.

The result of simulation shows that average collision rate is about 48 percent, average waste of transmission time is around 20.6 μs, and average utilization is about 91 percent. Figure 2 describes the relationship between collision rate and waste of transmission time. As the waste of transmission time increases, the average collision rate decreases inconsiderably. The average collision rate is so high that it can not be accepted in actual systems.

Case 2: with 64 ONUs, the deviation of RTT is allowed in range of [-2,2]. The results of 20 successive cycles of simulation are shown in table 2.

Table 2.  Results of the case of Δx = ± 2

| Cycle | ONU | Collision Rate | Waste of Trans Time (μs) | Line Utilization |
|-------|-----|----------------|--------------------------|------------------|
| 1 | 64 | 51.56 | 35.78 | 79.93 |
| 2 | 64 | 51.56 | 37.36 | 80.77 |
| 3 | 64 | 45.31 | 38.04 | 81.04 |
| 4 | 64 | 46.87 | 38.22 | 81.38 |
| 5 | 64 | 46.87 | 39.66 | 81.67 |
| 6 | 64 | 50.00 | 40.46 | 82.16 |
| 7 | 64 | 51.56 | 40.50 | 82.20 |
| 8 | 64 | 45.31 | 41.46 | 82.34 |
| 9 | 64 | 43.75 | 41.54 | 82.57 |
| 10 | 64 | 50.00 | 42.12 | 82.94 |
| 11 | 64 | 51.56 | 42.18 | 83.01 |
| 12 | 64 | 48.43 | 42.20 | 83.36 |
| 13 | 64 | 50.00 | 42.78 | 83.38 |
| 14 | 64 | 50.00 | 45.36 | 83.46 |
| 15 | 64 | 50.00 | 46.26 | 83.47 |
| 16 | 64 | 50.00 | 46.54 | 83.61 |
| 17 | 64 | 51.56 | 47.88 | 83.71 |
| 18 | 64 | 46.87 | 48.80 | 83.81 |
| 19 | 64 | 43.75 | 49.10 | 84.44 |
| 20 | 64 | 43.75 | 49.20 | 84.63 |

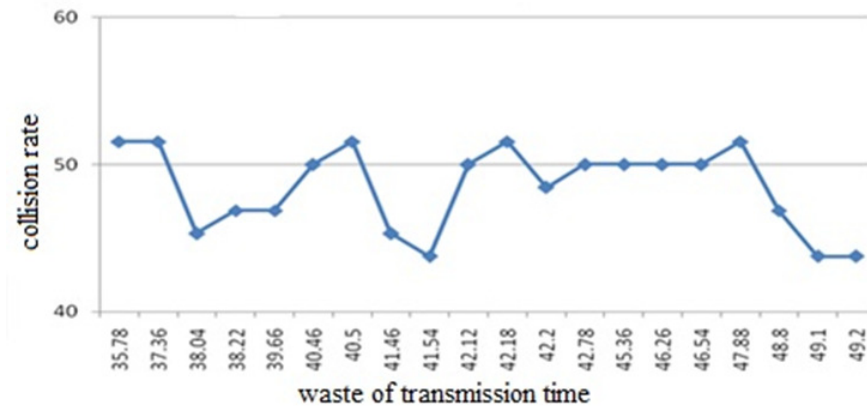

Figure 3. Relationship between collision rate and waste of transmission time with Δx = ± 2.

In the case of simulation, average collision rate is also 48 percent, but average waste of transmission time increases to 42.6 μs, and average line utilization decreases to 82 percent. Similar to case 1, figure 3 shows that average collision rate is still high, although the waste of transmission time increases. Both above cases of simulation show that the larger the RTT inaccuracy, the higher the collision rate and the smaller the line utilization.

The performance may seriously degrade due to RTT inaccuracy. As early discussed, the RTT inaccuracy frequently occurs in actual system by many reasons. It is necessary to have a solution that limits the collision rate and keeps the line utilization in high.

## 4. PROPOSED SOLUTION

As mentioned in the previous sections, if ONU(i) early starts to transfer data while ONU(i-1) has not finished transmission yet, a collision occurs (Δt <0). Conversely, if ONU(i) starts too late, it wastes its granted time (Δt >0). To combat the drawbacks due to RTT inaccuracy, the general idea is to make OLT to be aware of the possibility of RTT deviation in each ONUs. The deviation may be an offset between the expected value of RTT in the MPCPDU and the actual one. Based on this offset, OLT corrects the value of RTT in calculations. However, this heavy process could lead to more delay. Thus, it is necessary to consider a more feasible solution for the problem. This is one of our goals in this paper. Since the case of confliction is popular in real systems that crowded with a lot of ONUs, our solution aims to reduce the probability of collision. For the sake of that, we allow OLT to randomly take a time complement, say $C$, which is added to calculating RTT for assigning the start times for upstream bandwidth grants. By this way, ONU(i) is prone to soon starting with $C$, so avoiding overlapping on the start time of ONU(i+1). The range of $C$ depends on the possible deviation of RTT which is determined by carefully benchmarking real systems. In our simulation, we find that the suitable range of $C$ is a half of RTT deviation range. If RTT deviation changes in [-Δx; Δx], $C$ is in [0; Δx].

We examine our proposal by simulation with 64 ONUs, RTT deviation in [-1,1], and the complement $C$. The results of 8 successive cycles of simulation with complement $C$ are shown in table 3.

Table 3. Results of the case of Δx = ± 1 with complement *C*.

| Cycle | Collision Rate | Waste of Trans Time (µs) | Line Utilization |
|-------|----------------|--------------------------|------------------|
| 1 | 04.68 | 60.52 | 94.29 |
| 2 | 06.25 | 61.42 | 94.13 |
| 3 | 03.12 | 61.58 | 94.30 |
| 4 | 06.25 | 61.75 | 94.09 |
| 5 | 03.12 | 62.14 | 94.25 |
| 6 | 01.56 | 62.36 | 94.32 |
| 7 | 04.68 | 62.74 | 94.11 |
| 8 | 01.56 | 62.90 | 94.27 |

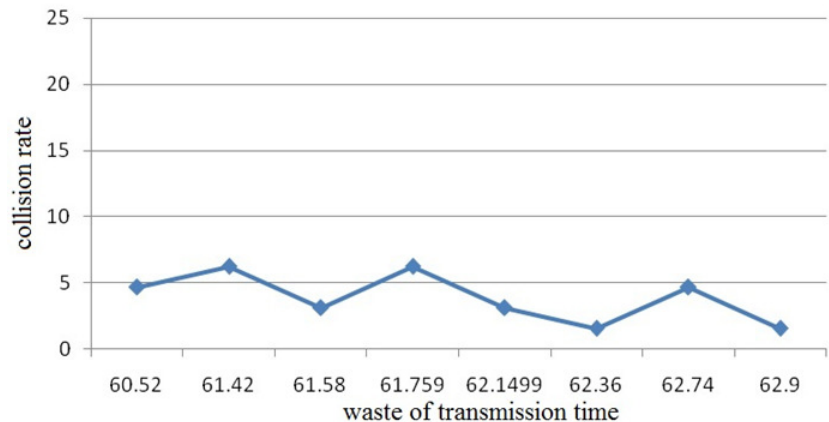

Figure 4. Relationship between collision rate and waste of transmission time with C.

The results show that average collision rate is reduced to 3.9 percent, the average waste of transmission time is about 62 µs, and the average line utilization increases to 94.2 percent. The collision rate decreases significantly in figure 4. These reasonable values result from adding the value of *C* to RTT. Indeed, *C* is a correction parameter that mitigates the RTT inaccuracy.

## 5. CONCLUSIONS

The impacts of RTT inaccuracy on DBA performance are analyzed in this paper. The simulation results show that the performance of DBA gets worse in conditions of RTT inaccuracy. Unfortunately, calculated RTT value frequently differs from its actual value by many reasons. Collision of adjacent transmissions is prone to high rate in the scheduling scheme of current DBAs due to the inaccuracy. The more collision rate the system gets, the more bandwidth the system wastes, and the less utilization the system earns. Our proposed method is a good prospect of combating the RTT inaccuracy. By proactive adding a complement to calculating RTT, our proposal reduces the probability of transmission collision between ONUs. The proposed method is validated by the results of simulation, which show that collision rate decreases to feasible value and the line utilization increase to high level. The way of mitigating RTT inaccuracy is also useful in related issues, such as determining the ONU sleep state period to reduce the energy consumption of PONs [3].

**Authors:**

**Son Nguyen Hong**, received his B.Sc. in Computer Engineering from University of Technology HCM city, his M.Sc. and PhD in Communication Engineering from the Post and Telecommunication Institute of Technology Hanoi in 2010. His current research interests include communication engineering, telecommunication networks, network security, computer engineering and cloud computing.

**Hao Nguyen  Anh**, received his B.Sc. in Computer Science from University of  Science HCM city, his M.Sc in Computer Science from AIT Thailand in 2003. His current research interests include communication and information systems, computer engineering and embedded systems.

**Thua Huynh Trong**, received his B.Sc in Computer Science from University of Science HCM city, his M.Sc in Communication Engineering from Kyung Hee University, South Korea in 2004. His current research interests include embedded systems, communication technology and wireless sensor networks. .